# Highly Efficient Waveform Design and Hybrid Duplex for Joint Communication and Sensing

Yihua Ma, Zhifeng Yuan, Shuqiang Xia, Guanghui Yu, and Liujun Hu

*Abstract*—**Joint communication and sensing (JCAS) is a very promising 6G technology, which attracts more and more research attention. Compared with communication, radar has many unique features in terms of waveform design criteria, self-interference cancellation (SIC), aperture-dependent resolution, and virtual aperture. This paper proposes a novel waveform design named max-aperture radar slicing (MaRS) to gain a large time-frequency aperture, which is generated by orthogonal frequency division multiplexing (OFDM) and occupies only a tiny fraction of OFDM resources. The proposed MaRS keeps the radar advantages of constant modulus, zero auto-correlation sequence, and simple SIC. As MaRS consumes much less resources, conventional processing methods fail, and novel angle-Doppler map based methods are proposed to obtain the range-velocity-angle information from MaRS echoes and strong clutters. To avoid complex full-duplex communication, this paper proposes a hybrid-duplex JCAS scheme composed of half-duplex communication and full-duplex radar. The half-duplex communication antenna array is reused, and a small sensing-dedicated antenna array is added. Using these two arrays, a large space-domain sensing aperture is virtually formed to greatly improve the angle resolution. The numerical results show that the proposed MaRS and hybrid duplex can achieve a high sensing resolution with only 0.4% OFDM resources, which reduces the overheads of conventional methods to less than one tenth.**

## I. INTRODUCTION

AS a popular 6G technology, joint communication and sensing (JCAS) [1]-[5] is expected to create considerable add-on values to the wireless communication system. The widely-deployed communication infrastructures can be enhanced to provide radar services like traffic control and surveillance, drone detection [6], and railway obstacle detection. JCAS can also be realized by the various mobile communication devices in the scenarios of autonomous driving [7][8], smart home, and health care.

This work was supported by the National Key Research and Development Program of China (No. 2021YFB2900200). *(Corresponding author: Yihua Ma.)*

The authors are with the State Key Laboratory of Mobile Network and Mobile Multimedia Technology, Shenzhen, China. They are also with the ZTE Corporation, Shenzhen, China. (e-mail: {yihua.ma, yuan.zhifeng, xia.shuqiang, yu.guanghui, hu.liujun}@zte.com.cn)

The communication and radar utilizes electromagnetic waves in two different ways. The idea of dual-function design can be traced back to the 1960s according to [2]. It attracts more and more research attention in recent years, and there are many driving factors including (1) the spectrum has been well exploited for two separate systems, and the joint spectrum utilization is expected to improve the efficiency and flexibility [2][9]; (2) the hardware designs have a similar technology trend of multiple antennas and digital baseband [4][10], and hardware sharing saves the cost; (3) the information fusion and mutual reinforcement of two functions can improve performance, especially for autonomous vehicles [7][8].

From the perspective of system integration, existing JCAS schemes are categorized into three levels: coexistence, cooperation, and joint design. In coexistence schemes, the two sub-systems transmit the signals in overlapped resources. Thus, it requires interference management to ensure the performance [11]. The cooperation of two functions helps to manage the interference via spectrum sharing [12] and beamforming [13][14]. The cognitive radar [15] can be seen as a special form of cooperation where the secondary function dynamically senses the spectrum and finds idle resources to transmit, whose idea is similar to that of cognitive radio [16]. To reduce spectrum sensing overheads, a sub-Nyquist radar can be used to sense the communication resources at low rates [17]. Unlike coexistence and cooperation, the joint design aims to integrate the two sub-systems into one, which requires the sharing of spectrum and hardware to some extend.

From the perspective of waveform design, existing works include the radar-centric, communication-centric and new waveform. The radar-centric waveform embeds data information in the radar signal. The most popular radar waveform is chirp [18] for its constant modulus, zero auto-correlation, and simple self-interference cancellation (SIC). Chirp spread spectrum (CSS) modulation has been used in Long Range Radio (LoRa) [19], and similarly, JCAS can utilize chirp to transmit information via phase and frequency modulation [20][21]. Apart from chirp radar, frequency-agile radar [22] was also utilized for JCAS which employs an index modulation of sub-carriers. Furthermore, the beam pattern of radar was proposed for communication via sidelobe control and waveform diversity [23]. A severe problem of radar-centric JCAS schemes is the limited data rate.

The communication-centric waveform uses communication signals to sense. Spread-spectrum communication was used in early times, and JCAS using phase-coded waveform [10] was proposed. Nowadays, orthogonal frequency division multiplexing (OFDM) has been widely used. In the OFDM-based passive radar [24][25], a passive node receives communication signals and realizes bi-static sensing. For widely used mono-static sensing scenarios, full duplex is required to realize an OFDM-based JCAS [26][27]. With full duplex, the base station processes OFDM echos for distance and Doppler estimation. Joint optimization methods can be used to make a good trade-off. In [28], the power allocation of each sub-system is optimized to maximize mutual information. The data symbol can be optimized to fill unused sub-carriers to minimize the estimation variances [29]. Furthermore, the waveform over multiple antennas can be optimized for both two sub-systems [30].

Apart from radar-centric and communication-centric schemes, some works [31][32] proposed to use brand-new JCAS waveforms. Orthogonal chirp division multiplexing (OCDM) [31] was proposed to replace the Fourier transform in OFDM with the Fresnel transform, which also has a DFT-spread-OFDM transceiving scheme [32] to generate similar waveform. Although OCDM methods modifies OFDM by chirp, the multi-path robustness of OFDM and efficient SIC of chirp can no longer be achieved. To keep both advantages, separate waveforms can be integrated [33][34]. The waiting time of pulse signal is used for OFDM communication in [33] to increase efficiency. [34] uses DFT spreading to multiplex chirp and data signal in one OFDM symbol and employs random sampling via adjusting the multiplexing position to improve sensing resolution.

Using the radar waveform achieves high sensing performance with simple SIC hardware. The limitation is the sensing resource overheads. Using existing OFDM signal to sense achieve a relatively worse performance but greatly saves the sensing resources. However, a severe problem of using existing OFDM signal is the dependence on in-band full-duplex (IBFD) communication. Although many IBFD technologies and prototypes [35][36] have been proposed, the benefit of double throughput is not that attractive as double antennas are required. With double antennas in half-duplex MIMO, a 2-fold throughput can easily be obtained without handling cumbersome self interference (SI). The full duplex of single antenna requires an ideal nonreciprocal ferrite circulator [36], which cannot be applied in MIMO as the circulator does not isolate SI across different antennas. In 3GPP Release 18 [37], an aggregation of two different sub-bands named sub-band full duplex, instead of IBFD, is in discussing.

Compared with communication systems, radar systems have unique advantages including (1) waveform superiority in terms of constant-amplitude zero auto-correlation sequence; (2) simple SIC in the analog domain; (3) aperture-dependent resolution instead of resource-dependent; (4) virtual aperture (VA) of multiple-input and multiple-output (MIMO) radar [38][39] to increase angle resolution. However, these advantages are hard to achieve using existing sensing waveforms and duplex schemes. The aim of this paper is to fully utilize these radar features via designing both the low-overhead aperture-based waveform and the efficient hybrid duplex scheme containing different duplex strategies for communication and sensing. The main contributions include:

1. This paper proposes a novel max-aperture radar slicing (MaRS) waveform to greatly reduce resource overheads without sacrificing radar resolution. Different structures of MaRS are proposed and compared, and comb MaRS performs best. Comb MARS keeps the advantages of constant modulus over time, zero auto-correlation, and very simple analog SIC.

2. As MaRS cannot be processed by conventional detection methods, this paper proposes novel angle-Doppler map (ADM) based methods to detect targets in ADM instead of range-Doppler map (RDM). Using this structure, a joint space-time target extraction and space-time adaptive processing (STTE-STAP) algorithm is proposed. Moreover, a novel space-time hierarchical processing (STHP) is proposed to avoid large matrix inverse and increase calculation accuracy.

3. A hybrid duplex JCAS scheme is first proposed, which comprises half-duplex communication and full-duplex radar. A small sensing dedicated antenna array is added to a typical communication system. It is combined with the communication array to form a large VA, which greatly enhances angle resolution. To simultaneously support MaRS and OFDM data transmission, a 2-step filter based receiver design is also proposed.

4. This paper provides simulations to verify the proposed schemes, which consider noise, SI, and environmental clutters. The simulation results show that high-resolution sensing is achieved with less than 0.4% resource overheads in an OFDM communication system, which means the communication can use almost 100% of resources to ensure throughput.

The rest of the paper is organized as follows. Section II shows the system model of communication, radar, and angle of arrival (AoA). Section III proposes several MaRS structures and ADM-based algorithms to recover sensing information from MaRS. Section IV proposes a hybrid duplex for JCAS which avoids IBFD communication and enables a large VA. Section V shows the simulation results of car and drone scenarios. Section VI briefly concludes the paper. In this paper, $(\cdot)^*$, $(\cdot)^T$, and $(\cdot)^H$ denote conjugate, transpose, and Hermitian transpose of a matrix or vector. $\otimes$, $\circ$ and mod represent the Kronecker product, Hadamard product, and modulo operation.

## II. SYSTEM MODEL

This paper considers a JCAS scenario containing cellular communication and mono-static sensing as shown in Fig. 1. A novel hybrid scheme is considered. The communication system is unchanged, which uses an half-wavelength-spacing array. This array also transmits the sensing signal, and a sensing-dedicated larger-spacing array receives the echoes.

### A. Communication Sub-system

The communication sub-system is a common half-duplex cellular system. In the downlink, the base station transmits an

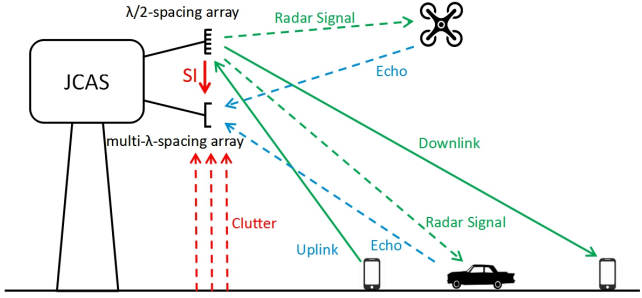

Fig. 1. The JCAS scenario considered in this paper.

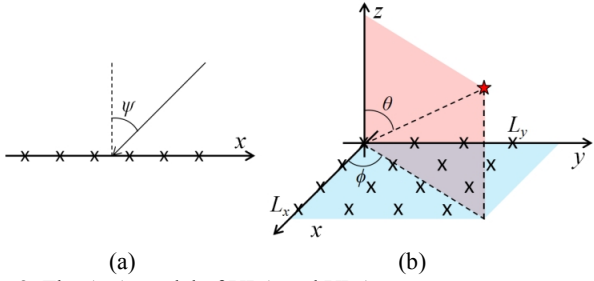

(a)          (b)

Fig. 2. The AoA model of ULA and URA.

OFDM signal of

$$x(t) = \frac{1}{\sqrt{N}} \sum_{m=0}^{M-1} rect(t - mT_{sym}) \sum_{n=0}^{N-1} X_{n,m} \cdot e^{j2\pi n\Delta f(t - mT_{sym})}, \quad (1)$$

where $N$, $M$, $\Delta f$, and $T_{sym}$ denote the number of sub-carriers, the number of symbols in a JCAS frame, the sub-carrier spacing, and the time duration of one symbol, respectively. Function $rect(t)$ is a time window of

$$rect(t) = \begin{cases} 1, & 0 \le t < T_{sym} \\ 0, & t < 0 \ or \ t \ge T_{sym} \end{cases}. \quad (2)$$

The cyclic prefix (CP) of OFDM is omitted for simplicity. The bandwidth of $x(t)$ is $B = N\Delta f$.

The transmit power is assumed to be $P_T$. The received power of the communication signal is

$$P_R = \frac{P_T G_T G_R \lambda^2}{(4\pi)^2 R^2}, \quad (3)$$

where $G_T$, $G_R$, $\lambda$, and $R$ denote the transmit gain, the receive gain, the wavelength, and the distance, respectively.

### B. Sensing Sub-system

This paper considers a radar scenario of simultaneously detecting multiple targets, or the searching mode. The sensing sub-system can reuse the OFDM baseband to generate radar signals. Chirp can be generated using OFDM symbols with a time-domain signal of

$$x_{chirp}(t) = e^{j\pi\mu t^2}, 0 \le t < T_{sym}, \quad (4)$$

where $\mu$ is the chirp rate. The frequency of this signal rises from 0 to $\mu T_{sym}$. To utilize the whole bandwidth of $x(t)$, the chirp rate is

$$\mu = \frac{B}{T_{sym}} = \frac{N\Delta f}{T_{sym}}. \quad (5)$$

The chirp signal is discretized as the time-domain OFDM symbol of

$$x_{chirp}[k] = e^{j\pi\mu(k/N \cdot T_{sym})^2}, 0 \le k < N. \quad (6)$$

It is generated by the frequency-domain OFDM symbols of

$$X_{chirp,n,m} = \frac{1}{\sqrt{N}} \sum_{k=0}^{N-1} x_{chirp}[k] \cdot e^{-jnk\frac{2\pi}{N}}. \quad (7)$$

The received power of the echo signal is

$$P_R' = \frac{P_T G_T G_R \lambda^2 \sigma}{(4\pi)^3 R^4}, \quad (8)$$

where $\sigma$ is the radar cross section of the target.

Apart from the echo signal of targets, the receiver also receives strong SI signal and clutter signal. SI is not only from scattering but also from near-field coupling, and the received SI signal in the $l$-th antenna is modeled as

$$y_l(t) = \sqrt{P_{SI}} e^{j\Omega_l} x(t), \quad (9)$$

where $P_{SI}$ and $\Omega_l$ is the power of SI and a random variable with a uniform distribution in $[0, 2\pi)$. Note that Equation (9) is a simplified model for multi-antenna SI as this paper uses simple radar self-interference cancellation methods, which is irrelevant to the phases at multiple antennas.

This paper also considers the environmental clutters. The Georgia Tech Research Institute (GTRI) model of the area reflectivity $\sigma^0$ [18] is used to describe the reflection of area scatters, which is given by

$$\sigma^0 = A(\delta + C)^B \exp\left(\frac{-D}{1 + 0.1\sigma_h/\lambda}\right), \quad (10)$$

where $\delta$ is the grazing angle, $\sigma_h$ is the RMS surface roughness, and the parameters $A$, $B$, $C$, and $D$ depend on the clutter type and radar frequency, which can be found in [40].

### C. AoA Model

As Fig. 2(a) shows, the steering vector of a uniform linear array (ULA) is

$$\mathbf{a}(\psi, L, d) = \begin{bmatrix} e^{j2\pi\frac{0 \cdot d \sin\psi}{\lambda}} & e^{j2\pi\frac{1 \cdot d \sin\psi}{\lambda}} & \dots & e^{j2\pi\frac{(L-1)d\sin\psi}{\lambda}} \end{bmatrix}^T, \quad (11)$$

where $\psi$, $L$ and $d$ denote the AoA, the antenna number and the spacing of antennas.

This paper considers an $L_x \times L_y$ uniform rectangular array (URA). As shown in Fig. 2(b), the elevation angle $\theta \in [0, \pi]$ and azimuth angle $\phi \in [-\pi, \pi)$ are used to decide the target direction. The direction vector from URA to the target is

$$\vec{a} = [\cos\phi\sin\theta, \sin\phi\sin\theta, \cos\theta]. \quad (12)$$

The AoA of URA at the x-y plane can also be modeled by the AoA of the x-axis ULA and the y-axis ULA, which are denoted by $\psi_x$ and $\psi_y \in [0, \pi]$. The direction vector becomes

$$\vec{a} = \left[\sin\psi_x, \sin\psi_y, \sqrt{1 - \sin^2\psi_x - \sin^2\psi_y}\right]. \quad (13)$$

Equations (12) and (13) can be converted to each other. This paper uses $\psi_x$ and $\psi_y$ as they can be directly estimated. The 2-dimension (2D) steering vector of a URA is

$$\mathbf{a}_{2D}(\psi_x, \psi_y, L_x, L_y, d_x, d_y) = \mathbf{a}(\psi_y, L_y, d_y) \otimes \mathbf{a}(\psi_x, L_x, d_x), \quad (14)$$

where $\otimes$ is the Kronecker product. As Equation (11) can be seen as a column vector of the discrete Fourier transform

(DFT) matrix $\mathbf{F}_L \in \mathbb{C}^{L \times L}$. The 2D vector in Equation (14) can be seen as a column vector of a synthesis FFT matrix of

$$\mathbf{F}_{L_x, L_y} = \mathbf{F}_{L_y} \otimes \mathbf{F}_{L_x} . \tag{15}$$

## III. MaRS Structures and Processing Methods

### A. Sensing Aperture Analysis

The sensing function requires a large aperture in the frequency, time, and space domain to ensure the resolution of range, velocity, and angle, respectively. The range resolution is decided by the bandwidth $B$ as

$$\Delta R = \frac{c}{2B} . \tag{16}$$

A coherent processing interval (CPI) of $T_{\text{CPI}} = MT_{sym}$ is assumed, and the velocity resolution is

$$\Delta v = \frac{c}{2T_{\text{CPI}} f_c} . \tag{17}$$

The space-domain aperture $D = Ld$ decides the angle resolution, which can be calculated by

$$\Delta \psi \approx \alpha \frac{\lambda}{D} \text{ rad} . \tag{18}$$

The coefficient $\alpha$ is a beamwidth-related factor [40]. To get rid of this, this paper uses the resolution of $\sin\psi$ as

$$\Delta \sin \psi = \frac{\lambda}{D} . \tag{19}$$

The existing methods includes two kinds. The first is to use a dedicated wideband sensing signal to achieve a large aperture. The problem is that a great deal of resources are required to ensure sensing performance. The second is to optimize a joint waveform for both communication and sensing to save resources. The problem is the complexity and delay of optimization and the performance gain degradation for non-overlapped sensing and communication targets. This paper proposes a third way of gaining a large sensing aperture with only a tiny portion of resources for sensing.

### B. Basic MaRS Structures

Based on the analysis of sensing aperture, high-resolution radar sensing requires a large aperture, instead of a large number of resources. If a large aperture is realized by few resource overheads, high sensing resolution is still possible to achieve. Then, the majority of time-frequency resources can be used for communication to ensure the data rate. Therefore, this paper proposes a sensing waveform design named MaRS. Note that MaRS not only acts as a resource allocation strategy, but also involves with the transmitted signal design.

Fig. 3(a) shows the OFDM resource overheads to generate the conventional wideband chirp signal, which consumes all the resources in the time-frequency-space aperture. Here, continuous OFDM symbols are used for simplicity. The time-frequency domain resource overhead is measured by the number of resource elements (RE) in OFDM. The space-domain resources can be measured by the number of antennas, or the space-domain degree of freedom.

A time-frequency-space MaRS signal is shown in Fig. 3(b).

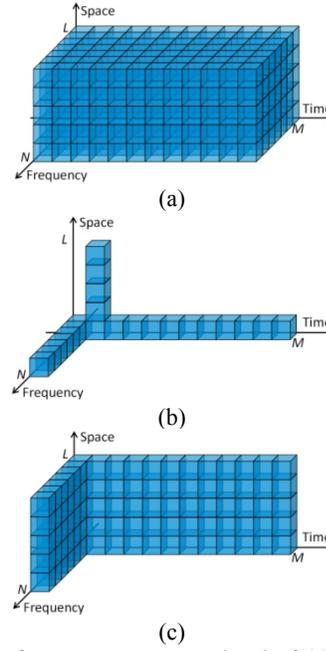

Fig. 3. The time-frequency-space overhead of (a) the conventional chirp signal, (b) time-frequency-space MaRS signal and (c) time-frequency MaRS signal.

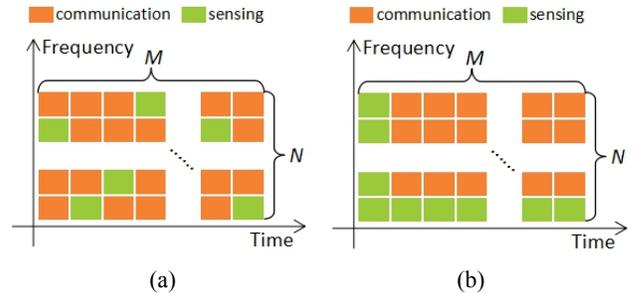

Fig. 4. Two examples of time-frequency MaRS signal as (a) frequency-agile MaRS and (b) L-shape MaRS.

The transmission of such a signal can be divided into three stages: (1) in the first symbol, the JCAS station transmits the chirp signal using $N$ sub-carriers and uses only one pair of transmit antenna and receive antennas; (2) in the second symbol, the JCAS station transmits the single-tone sensing signal and uses all the antennas; and (3) in the left ($M$-2) symbols, the JCAS station transmits the single-tone sensing signals and uses only one pair of transmit antenna and receive antennas. The phase of the single-tone part of MaRS is set to be continuous for the extraction at the receiver. The resolution is as high as that of conventional schemes in Fig. 3(a) except that the radial velocity resolution is slightly reduced as the useful time-domain aperture is reduced from $MT_{sym}$ to ($M$-2)$T_{sym}$. However, the association of estimated range, velocity, and angle is hard to decide, which makes it only work in the ideal single-target and clutter-free scenario. Also, the remaining space-domain resources cannot be used by communication as its SI is hard to be removed.

To avoid these problems of time-frequency-space MaRS, a time-frequency MaRS signal is proposed as in Fig. 3(c), and two structures of time-frequency MaRS are shown in Fig. 4.

The frequency-agile (FA) MaRS in Fig. 4(a) transmits a random sub-carrier in each OFDM symbol, which is similar to the sensing part of [22]. The communication part is different as the left OFDM resources are for data transmission to gain a high data rate. However, the sensing performance of FA MaRS is limited. L-shape MaRS is proposed to use dedicated resources for range and velocity estimation as shown in Fig. 4(b). It is divided into two stages: (1) in the first symbol, the JCAS station transmits the chirp signal using $N$ sub-carriers; (2) in the left ($M$-1) symbols, the JCAS station transmits a continuous-phase single-tone sensing signal. The resource overhead is greatly reduced from $MN$ REs to ($M$+$N$-1) REs.

## C. ADM-based Method

MaRS signal cannot be processed by conventional radar methods. The radio-frequency (RF) chirp signal is

$$x_c(t) = e^{j2\pi f_c t} \sum_{m=0}^{M-1} rect(t - mT_{sym}) e^{j\pi\mu(t-mT_{sym})^2}, 0 \le t < MT_{sym}. \tag{20}$$

The echo signal caused by the scattering of a point target at a distance of $c\tau/2$ is

$$y_{echo}(t) = \mathbf{a}_{2D}(\psi_x, \psi_y, L_x, L_y, d_x, d_y) h_0 x_c(t-\tau) e^{j2\pi f_d(t-\tau)}, \\ \tau \le t < MT_{sym} + \tau, \tag{21}$$

where $h_0 \in \mathbb{C}$ is the round-trip channel coefficient in the symbol with index 0.

The chirp receiver mix each row of $\mathbf{y}_{echo}(t)$ with $x_c(t)$ via the mixer, and the intermediate frequency (IF) signal is

$$\mathbf{y}_{IF}(t) = \mathbf{a}_{2D}(\psi_x, \psi_y, L_x, L_y, d_x, d_y) h_0 e^{j2\pi\mu\tau t} e^{j2\pi f_d(t-\tau)}, \\ mT_{sym} + \tau \le t < (m+1)T_{sym} + \tau, 0 \le m < M. \tag{22}$$

As the use of OFDM assumes $f_D T_{sym} \ll 1$, $f_D t \ll f_D T_{sym} \ll 1$ can be omitted. With $\tau = 0$, the IF SI signal is a direct-currency (DC) signal, which can be removed via a DC offset cancellation (DCOC) circuit [41]. Further enhancement can be added to DCOC to make the transition band decay faster [42].

The receiving window is set as [$mT_{sym}$, ($m$+1)$T_{sym}$), and the signals out of the receiving window are omitted. The continuous-time IF signal of $M$ symbols and $L$ antennas after filtering can be sampled to form a matrix $\mathbf{Y}_F \in \mathbb{C}^{L \times MN}$. $\mathbf{Y}_F$ can be processed via range, Doppler and angle FFT. The space-domain $L$-FFT in this figure can be replaced by Equation (15) when URA is used. The moving targets can be detected in the RDM of each angle [6], and these operations can be seen as a 3D-FFT with a total complexity of O($LMN$log($LMN$)). As a comparison, MaRS does not have so many REs to implement 3D-FFT and construct RDM.

To extract moving targets from low-overhead MaRS signal,

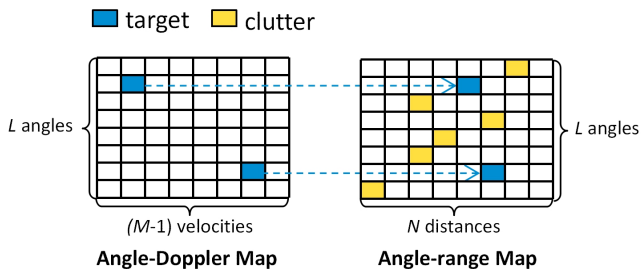

Fig. 5. The ADM-based method of L-shape MaRS signal.

an ADM-based method is proposed as shown in Fig. 5. The RF L-shape MaRS signal is

$$x_m(t) = e^{j2\pi f_c t} rect(t) e^{j\pi\mu t^2} \\ + e^{j2\pi f_c t} \sum_{m=1}^{M-1} rect(t - mT_{sym}) e^{j2\pi f_{ST} t}, 0 \le t < MT_{sym}, \tag{23}$$

where $f_{ST}$ is the single-tone sub-carrier frequency. By replacing $x_c(t)$ with $x_m(t)$ in Equation (21), $\mathbf{y}'_{echo}(t)$ can be derived, and the IF signal is

$$\mathbf{y}'_{IF}(t) = \begin{cases} \mathbf{y}'_{IF1}(t) = \mathbf{a}_{2D}(\alpha, \beta) h_0 e^{j2\pi\mu\tau t} e^{j2\pi f_d(t-\tau)}, \\ \quad \text{for } \tau \le t < T_{sym} + \tau; \\ \mathbf{y}'_{IF2}(t) = \mathbf{a}_{2D}(\alpha, \beta) h_0 e^{j2\pi f_d(t-\tau)}, \\ \quad \text{for } mT_{sym} + \tau \le t < (m+1)T_{sym} + \tau, m \in [1, M). \end{cases} \tag{24}$$

The IF signal has two parts. The first part $\mathbf{y}'_{IF1}(t)$ is the same as that in the conventional chirp scheme, and the SI can be easily removed via DCOC. In the second part $\mathbf{y}'_{IF2}(t)$, the SI of sensing signal as well as the scattering point with $f_d = 0$ are becoming DC signal which can also be removed. $\mathbf{y}'_{IF2}(t)$ also suffers from the SI of the communication signal at other sub-carriers, and the receiver separates them via filtering. In this paper, the Butterworth filter is assumed as

$$|H(f)|^2 = \frac{1}{1 + (f/f_{cut})^{2a}}, \tag{25}$$

where $f_{cut}$ is the cutoff frequency, and $a$ is the order of Butterworth filter. In practice, the guard band can also be inserted between the single-tone sensing signal and the communications signal to improve the separation.

These two parts of the IF signal after filtering is sampled as $\mathbf{Y}'_{F1} \in \mathbb{C}^{L \times N}$ and $\mathbf{Y}'_{F2} \in \mathbb{C}^{L \times (M-1)N}$. 2D-FFT can be applied to these two matrices. Before 2D-FFT, $\mathbf{Y}'_{F2}$ is pre-processed as

$$\mathbf{Y}'_{S2} = \left[ \sum_{n=0}^{N-1} \mathbf{y}'_{F2,n} \quad \sum_{n=N}^{2N-1} \mathbf{y}'_{F2,n} \quad \cdots \quad \sum_{n=(M-2)N}^{(M-1)N-1} \mathbf{y}'_{F2,n} \right] \in \mathbb{C}^{L \times (M-1)}, \tag{26}$$

where $\mathbf{y}'_{F2,n}$ is the $n$-th column vector of $\mathbf{Y}'_{F2}$, $n = 0, 1, 2, .., N$-1. This step reduces the data dimension, which not only reduces the complexity but also saves the storage. Applying 2D-FFT to $\mathbf{Y}'_{F1}$ and $\mathbf{Y}'_{F2S}$ as

$$\mathbf{R}'_1 = \mathbf{F}_{L_x, L_y} \left( \mathbf{F}_N \mathbf{Y}'^T_{F1} \right)^T = \mathbf{F}_{L_x, L_y} \mathbf{Y}'_{F1} \mathbf{F}_N. \tag{27}$$

$$\mathbf{R}'_2 = \mathbf{F}_{L_x, L_y} \left( \mathbf{F}_{M-1} \mathbf{Y}'^T_{S2} \right)^T = \mathbf{F}_{L_x, L_y} \mathbf{Y}'_{S2} \mathbf{F}_{M-1}. \tag{28}$$

With these two matrices, this ADM-based method can be realized to process L-shape MaRS as Fig. 5. The angle-velocity matrix $\mathbf{R}'_2$ is used to get the angle and velocity estimations of targets. Using the obtained angle estimation, the distance information is then obtained in the angle-distance matrix $\mathbf{R}'_2$. The complexity of 2-time 2D-FFT in this ADM-based method, O($LN$log($LN$) + $LM$log($LM$)), is also much less than that of the conventional 3D-FFT, O($LMN$log($LMN$)).

## D. Comb-MaRS and Space-time Processing

In the L-shape MaRS, the target distance information cannot be achieved when the clutters are strong and abundant, e.g., the rank of clutters in $\mathbf{Y}'_{F1}$ is $L$, and meanwhile targets are weaker than the clutters in the corresponding angles. To solve this problem, further enhancements are proposed.

An enhanced structure named comb MaRS is shown in Fig. 6(a). It transmits the wideband chirp signal in several OFDM symbols. The comb MaRS signal is

$$x_{cm}(t) = e^{j2\pi f_c t} \sum_{m=i_0}^{i_{w-1}} rect(t - mT_{sym}) e^{j\pi \mu (t - mT_{sym})^2}$$

$$+ e^{j2\pi f_c t} \sum_{m=0\sim M-1}^{m \neq i_w} rect(t - mT_{sym}) e^{j2\pi f_{ST} t}, 0 \leq t < MT_{sym}, \quad (29)$$

where $i_w \in [0, M-1]$ with $w = 0, 1, ..., W-1$ is the symbol index of the wideband chirp signal. The comb MaRS signal is a generalized version of the basic MaRS signal ($W = 1$).

This ADM-based procedure for comb MaRS is shown in Fig. 6(b). Comb MaRS enables STAP after target detection and angle-velocity estimation in ADM, and STTE-STAP is proposed. Similar to Equations from (21) to (24), the echo signal of $x_{cm}(t)$, $y''_{echo}(t)$, is obtained and mixed with $x_{cm}(t)$ to

$$y_{IF}''(t) = \begin{cases} y_{IF1}''(t) = \mathbf{a}_{2D}(\psi_x, \psi_y, L_x, L_y, d_x, d_y) h_0 e^{j2\pi \mu t (t - i_w T_{sym})} e^{j2\pi f_c (t-\tau)}, \\ \quad \text{for } i_w T_{sym} + \tau \leq t < (i_w + 1)T_{sym} + \tau, w \in [0, W); \\ y_{IF2}''(t) = \mathbf{a}_{2D}(\psi_x, \psi_y, L_x, L_y, d_x, d_y) h_0 e^{j2\pi f_c (t-\tau)}, \\ \quad \text{for } mT_{sym} + \tau \leq t < (m+1)T_{sym} + \tau, m \in [0, M) \ \& \ m \neq i_w. \end{cases}$$
$$(30)$$

These two parts of the IF signal after filtering are then sampled using a frequency of $1/B$ to obtain $\mathbf{Y}''_{F1} \in \mathbb{C}^{L \times WN}$ and $\mathbf{Y}''_{F2} \in \mathbb{C}^{L \times (M-W)N}$. The second part $\mathbf{Y}''_{F2}$ is transformed into $\mathbf{Y}''_{S2} \in \mathbb{C}^{L \times (M-W)}$ using the same operation of Equation (26). Unlike $\mathbf{Y}'_{S2}$ in Equation (26), $\mathbf{Y}''_{S2}$ is discontinuous at arbitrary $i_w$ in $[0, M-1]$. This discontinuous matrix is expanded to $\mathbf{Y}''_{E2}$

$\in \mathbb{C}^{L \times M}$ via the simple nearest-neighbour interpolation method. The $m$-th column vector of $\mathbf{y}''_{E2}$ is

$$\mathbf{y}''_{E2,m} = \begin{cases} \mathbf{y}''_{S2, m+g(m)-1}, m = i_w; \\ \mathbf{y}''_{S2, m+g(m)}, m \neq i_w, \end{cases} \quad (31)$$

where function $g(m)$ returns the number of $i_w$ values satisfying $i_w < m$. This equation requires $i_w$ to be non-zero. With $\mathbf{Y}''_{E2}$, ADM is obtained as

$$\mathbf{R}_2^n = \mathbf{F}_{2D}(L_x, L_y)(\mathbf{F}_M \mathbf{Y}''_{E2}{}^T)^T = \mathbf{F}_{2D}(L_x, L_y) \mathbf{Y}''_{E2} \mathbf{F}_M. \quad (32)$$

In ADM, STTE can be done, and $K$ targets are extracted with their velocity and angle estimations $\hat{v}_k, \hat{\psi}_{x,k}$ and $\hat{\psi}_{y,k}$. FFT in this estimation can be replaced by super-resolution estimation methods like MUSIC, which will be applied in simulations.

With the estimations from STTE, STAP can be realized. $\mathbf{Y}''_{F1}$ is reshaped to a space-time signal of

$$\mathbf{Y}_{ST} = \begin{bmatrix} \mathbf{y}''_{F1,0} & \mathbf{y}''_{F1,1} & \cdots & \mathbf{y}''_{F1,N-1} \\ \mathbf{y}''_{F1,N} & \mathbf{y}''_{F1,N+1} & \cdots & \mathbf{y}''_{F1,2N-1} \\ \cdots & \cdots & \cdots & \cdots \\ \mathbf{y}''_{F1,(W-1)N} & \mathbf{y}''_{F1,(W-1)N+1} & \cdots & \mathbf{y}''_{F1,WN-1} \end{bmatrix} \in \mathbb{C}^{WL \times N}. \quad (33)$$

The rank of clutter correlation in STAP [18] is

$$rank\{\mathbf{S}_C\} \approx L + (M-1)\beta, \quad (34)$$

where $\beta$ is clutter ridge slope, which is proportional to the speed of the radar platform. As the JCAS base station is static, $\beta = 0$. The space-time steering vector is constructed as

$$\hat{\mathbf{a}}_{ST,k} = \hat{\mathbf{a}}_{f_D,k} \otimes \mathbf{a}_{2D}(\hat{\psi}_{x,k}, \hat{\psi}_{y,k}, L_x, L_y, d_x, d_y) \in \mathbb{C}^{WL \times 1}, \quad (35)$$

where $\hat{\mathbf{a}}_{f_D,k} = \begin{bmatrix} e^{j2\pi \frac{\hat{v}_k f_{c_{i_0}}}{c}} & e^{j2\pi \frac{\hat{v}_k f_{c_{i_1}}}{c}} & \cdots & e^{j2\pi \frac{\hat{v}_k f_{c_{i_{W-1}}}}{c}} \end{bmatrix}^T \in \mathbb{C}^{W \times 1}$.

The interference correlation matrix is used to suppress the clutter and multi-target interference, which is estimated by

$$\hat{\mathbf{S}}_I = (\mathbf{Y}_{ST}^H \mathbf{Y}_{ST})/N. \quad (36)$$

The distance spectrum can be obtained using STAP together with a distance-dimension FFT as

$$\mathbf{d}_k = \frac{\left| (\hat{\mathbf{a}}_{ST,k}^H (\hat{\mathbf{S}}_I^{-1})^H (\mathbf{Y}_{ST} \mathbf{F}_N))^T \right|^2}{\hat{\mathbf{a}}_{ST,k}^H (\hat{\mathbf{S}}_I^{-1})^H \hat{\mathbf{a}}_{ST,k}} \in \mathbb{C}^{N \times 1}, \quad (37)$$

where $\mathbf{d}_k$ is a vector representing $N$ distance bins, and the bin of the largest value is the estimation of target range $R$. Now, the distance, velocity and angle information of multiple targets have all been obtained. $\hat{\mathbf{S}}_I$ is assumed to be full-rank with the presence of noise as it is easy to have $WL \leq N$. When $WL > N$, this method also works by using the pseudo-inverse.

### E. STAP analysis and Novel STHP

To analyse STAP, the eigen-decomposition of $\hat{\mathbf{S}}_I$ is made as

$$\hat{\mathbf{S}}_I = \mathbf{X} \begin{bmatrix} \lambda_1 & & & \\ & \lambda_2 & & \\ & & \ddots & \\ & & & \lambda_{WL} \end{bmatrix} \mathbf{X}^T. \quad (38)$$

Without loss of generality, the eigenvalues are assumed to be descending. The eigenvalues of $\lambda_1$ to $\lambda_{K+L}$ are from the target and clutters which is proportional to the signal power, while the eigenvalues of $\lambda_{K+L+1}$ to $\lambda_{WL}$ are from the noise which is proportional to the noise power.

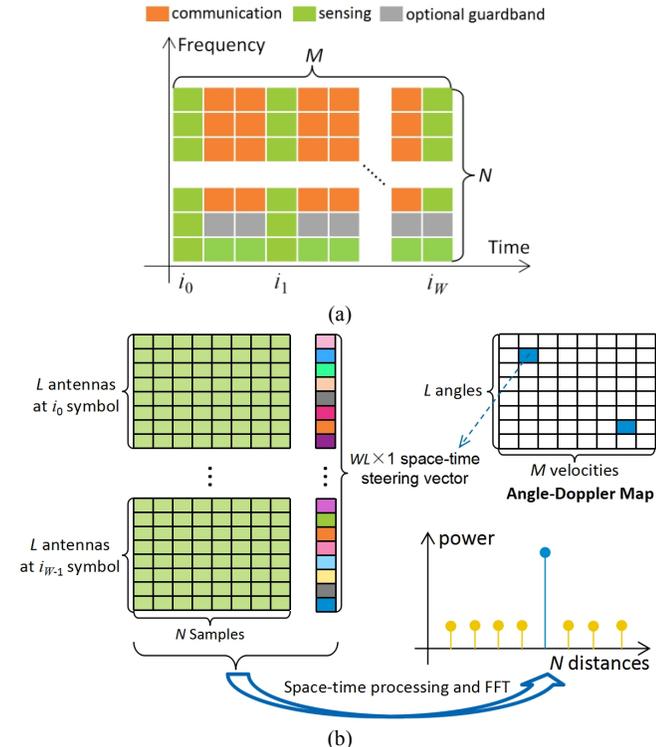

Fig. 6. (a) The time-frequency representation and (b) the ADM-based method of comb MaRS signal.

One problem is that when the signal-to-noise ratio (SNR) is higher, the condition number of $\hat{\mathbf{S}}_I$, $\lambda_1/\lambda_{WL}$, is larger, which means $\hat{\mathbf{S}}_I$ is more ill-conditioned. Note that $\hat{\mathbf{a}}_{ST,k}$ contains two parts, the true value and estimation errors. The true value matches the eigenvectors from the signal, while estimation errors cannot. Then, the inverse of $\hat{\mathbf{S}}_I$ amplifies the estimation errors of $\hat{\mathbf{a}}_{ST,k}$. Such a noise amplification won't happen if $\hat{\mathbf{a}}_{ST,k}$ is used to construct $\hat{\mathbf{S}}_I$ as the eigenvectors match $\hat{\mathbf{a}}_{ST,k}$. However, such calculations cannot be done as the space-time information of environmental clutters is hard to obtain. Therefore, there is a mismatch of $\hat{\mathbf{a}}_{ST,k}$ and $\hat{\mathbf{S}}_I$ in STTE-STAP.

To alleviate this mismatch, a noise-like item with a power of $\varepsilon$ can be added to improve the small eigenvalues as

$$\mathbf{d}_k = \frac{\left|\left(\hat{\mathbf{a}}_{ST,k}^H\left(\mathbf{R}_1 + \varepsilon\mathbf{I}_{WL}\right)^{-1}\left(\mathbf{Y}_{ST}\mathbf{F}_N\right)\right)^T\right|^2}{\hat{\mathbf{a}}_{ST,k}^H\left(\mathbf{R}_1 + \varepsilon\mathbf{I}_{WL}\right)^{-1}\hat{\mathbf{a}}_{ST,k}} \in \mathbb{C}^{N\times 1}, \quad (39)$$

where $\mathbf{I}_N$ is an $N\times N$ unit matrix. To make $\varepsilon$ effective to affect $\hat{\mathbf{S}}_I$, it is set to

$$\varepsilon_M = \frac{1}{WL} tr(\mathbf{R}_1), \quad (40)$$

where $tr(\cdot)$ is the trace of a matrix. In this case, the complexity becomes $O(LM\log(LM) + W^2L^2N + KW^2L^2 + W^3L^3 + KWLN)$.

When $\varepsilon\to\infty$, the matrix inverse can be removed and Equation (39) degrades to a matched filtering as

$$\mathbf{d}_k = \frac{\left|\left(\mathbf{a}_{ST,k}^H\left(\mathbf{Y}_{ST}\mathbf{F}_N\right)\right)^T\right|^2}{\mathbf{a}_{ST,k}^H\mathbf{a}_{ST}} \in \mathbb{C}^{N\times 1}. \quad (41)$$

The complexity is also reduced to $O(LM\log(LM)+KWLN)$.

This paper further proposes another method named space-time hierarchical processing (STHP) to alleviate $\mathbf{a}_{ST,k}/\hat{\mathbf{S}}_I$ mismatch and matrix inverse complexity. The space-time vector $\mathbf{a}_{ST,k}$ is divided into $\mathbf{a}_{T,k}\otimes\mathbf{a}_{S,k}$. In the $W$-dimension time domain, zero-forcing (ZF) is used to suppress the static clutter using the combining weight vector of

$$\mathbf{w}_{T,k} = \boldsymbol{\tau}_1, \boldsymbol{\Gamma} = \begin{bmatrix}\boldsymbol{\tau}_1\\\boldsymbol{\tau}_2\end{bmatrix} = \begin{bmatrix}\mathbf{a}_{T,k} & \mathbf{1}_{W\times 1}\end{bmatrix}^+, \quad (42)$$

where $\mathbf{1}_{W\times 1}$ is a $W\times 1$ all-one vector, and $(\cdot)^+$ is the pseudo-inverse. It combines $\mathbf{Y}_{ST}$ into $\mathbf{Y}_S \in \mathbb{C}^{L\times N}$. After the time-domain combination, the static clutter is forced to zero. Then, a matched filtering or adaptive processing can be done using

$$\mathbf{w}_{S,k} = \frac{\mathbf{a}_{S,k}^H\left(\mathbf{R}_1' + \varepsilon\mathbf{I}_L\right)^{-1}}{\sqrt{\mathbf{a}_{S,k}^H\left(\mathbf{R}_1' + \varepsilon\mathbf{I}_L\right)^{-1}\mathbf{a}_{S,k}}}, \mathbf{R}_1' = \left(\mathbf{Y}_S^H\mathbf{Y}_S\right)/N. \quad (43)$$

When $\varepsilon = 0$, it is an adaptive processing, while when $\varepsilon\to\infty$, it is a matched filter. A trade-off can also be made for $\varepsilon = \varepsilon_M$. The distance detection result is $\mathbf{d}_k = \mathbf{w}_{S,k}\mathbf{Y}_S$. Compared to STAP, the proposed STHP filters clutters by W-dimension ZF and then separates multiple targets by $L$-dimension processing. The clutter suppression performance can be ensured in STHP, and the matrix inverse complexity is reduced from $W^3L^3$ to $L^3$. The total complexity of STTE-STHP is $O(LM\log(LM) + L^2N + KL^2 + L^3 + KLN)$, and it decreases to $O(LM\log(LM) + KLN)$ when $\varepsilon\to\infty$.

## IV. Hybrid Duplex and Virtual Aperture

Section III shows how to achieve a high range-Doppler resolution with limited spectrum resources, and this section will explain how to gain a high angle-domain resolution with limited antennas.

### A. Hybrid Duplex Scheme

Full duplex, or SIC, is a must for mono-static sensing, while the communication has employed half-duplex schemes for decades. The IBFD communication usually requires a joint effort of antenna, analog and digital SIC as well as 2-fold antenna overheads. Also, the IBFD gain degrades greatly with the path loss and can even disappear, as SIC becomes harder for low-SNR users [43]. Therefore, sub-band full duplex is more practical [37], which cannot be used for sensing.

Unlike SIC of the variable communication data signal, the SI of sensing signal is simpler, and an efficient SIC can be achieved. Pulse radar uses a shared antenna to transmit and receive via switching, while chirp radar simply filters DC SI after mixing the received signal with the local chirp signal. It is easy for radar to achieve full duplex. Basing on the SIC realization difference of communication and radar, this paper proposes a novel hybrid-duplex JCAS scheme of half-duplex communication and full-duplex radar as shown in Fig. 7.

The proposed hybrid-duplex JCAS adds a small antenna array and the corresponding RF chain to realize full-duplex radar. Filters are required to realize the analog SIC for radar sensing. Also, a radar processing baseband is required, which can be implemented in a shared processing unit with the half-duplex OFDM baseband. The mixer in the RF chain of the large communication antenna array switches to the mixing of the local chirp signal in the $i_w$-th OFDM symbol.

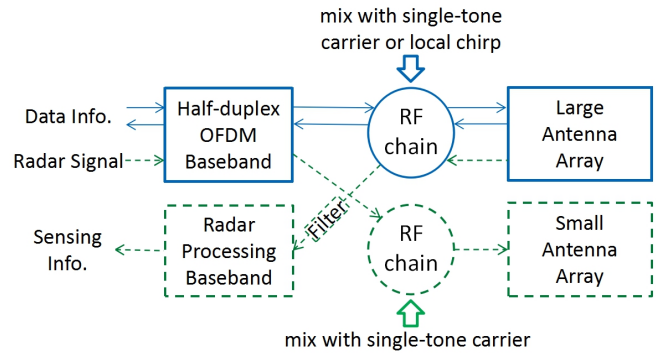

Fig. 7. The proposed hybrid duplex scheme. The blue solid lines represent the modules that has already existed in half-duplex communication systems. The green dashed lines represent the extra modules to realize a JCAS system.

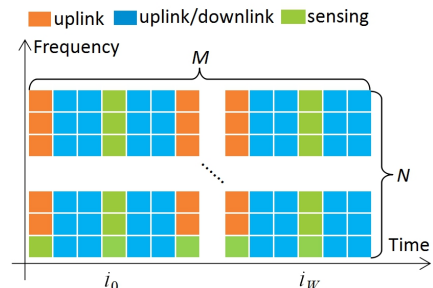

Fig. 8. The time-frequency representation of comb MaRS signal with a 1/3 duty ratio.

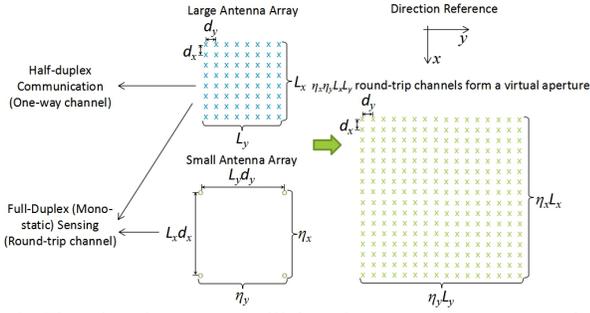

Fig. 9. The virtual aperture utilizing the two antenna arrays in the proposed hybrid-duplex JCAS scheme.

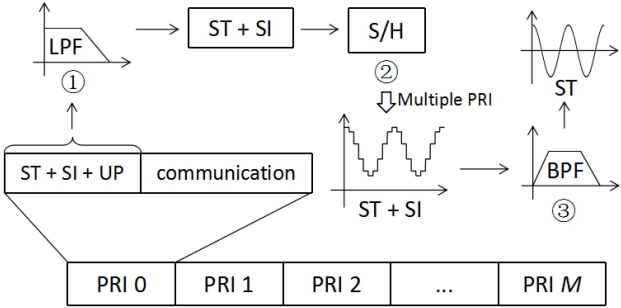

Fig. 10. The 2-step filter to processing the single-tone sensing signal. ST, SI and UP denote single-tone signal, self-interference, and uplink communication signal. One target is assumed for explanation.

Note that in the proposed scheme, the antenna cannot simultaneously transmit and receive. When the large antenna array is receiving the single-tone sensing signal, the downlink transmission is not supported. Fortunately, a 100% duty ratio MaRS signal is usually unnecessary in practice. For example, the MaRS signal with a 1/3 duty ratio is shown in Fig. 8. In the communication-only symbols, both uplink and downlink are supported, and the resource elements are marked blue in this figure. In the symbols containing single-tone sensing signals, only uplink is supported, and the resource elements are marked blue.

### B. Virtual Aperture in Hybrid Duplex

The proposed hybrid structure can be further utilized to realize VA, and the small sensing-dedicated array is customized. VA is a crucial technology to increase the angle resolution of MIMO radar [44]. VA utilizes the feature of the round-trip channel, and both the transmitting and receiving antennas provides phase differences. Therefore, $L_x$ transmitting antennas and $L_y$ receiving antennas can be utilized to provide a VA of $L_xL_y$ antennas. The spacing of the small array is designed to fit the large antenna array, and a uniform VA is formed as shown in Fig. 9. The antenna spacing of the small array is $L_xd_x$ and $L_yd_y$ in the column and row direction.

VA is a unique feature of the sensing function as the sensing signal experiences the round-trip channel. The sensing transmitting steering vector is

$$\mathbf{a}_T = \mathbf{a}(\psi_y, \eta_y, L_y d_y) \otimes \mathbf{a}(\psi_x, \eta_x, L_x d_x),\quad (44)$$

while the sensing receiving steering vector is

$$\mathbf{a}_R = \mathbf{a}(\psi_y, L_y, \lambda/2) \otimes \mathbf{a}(\psi_x, L_x, \lambda/2).\quad (45)$$

The channel information of total $\eta_x\eta_yL_xL_y$ MIMO channels forms a steering vector of large VA as

$$\begin{aligned}\mathbf{a}_{VA} &= \left(\mathbf{a}(\psi_y, \eta_y, L_y d_y) \otimes \mathbf{a}(\psi_y, \eta_y, d_y)\right) \otimes \\ &\quad \left(\mathbf{a}(\psi_x, \eta_x, L_x d_x) \otimes \mathbf{a}(\psi_x, \eta_x, d_x)\right) \\ &= \mathbf{a}(\psi_y, \eta_y L_y, d_y) \otimes \mathbf{a}(\psi_x, \eta_x L_x, d_x),\end{aligned}\quad (46)$$

which is equivalent to a URA of $\eta_x L_x \times \eta_y L_y$ antennas with the array spacing $d_x$ and $d_y$ in Fig. 9. Using VA, the aperture size can be greatly improved to $\eta_x\eta_y$ times.

The construction of VA requires every antenna in transmit array to send orthogonal signals. There are three ways to send orthogonal chirp signals, including time division [44], orthogonal code spreading [38][44], and cyclic shift [31][32]. The first two methods can also be used for the single-tone signal. Further, these methods can be used within an OFDM symbol or inter OFDM symbols. This paper employs the inter-symbol time-division method for example. The received sensing signal in these symbols is

$$\mathbf{Y}_O = \left[\mathbf{Y}_{1,1},...,\mathbf{Y}_{\eta_x,1},\mathbf{Y}_{1,2},...,\mathbf{Y}_{\eta_x,2},.......,\mathbf{Y}_{\eta_x,\eta_y}\right] \in \mathbb{C}^{L_x L_y \times \eta_x \eta_y N},\quad (47)$$

where $\mathbf{Y}_{a,b}$ denotes the OFDM symbol transmitted by the antenna $(a, b)$ in the small antenna array and received by the large antenna array. The $\eta_x\eta_y$ symbols in the $\mathbf{Y}_O$ can be converted to a VA receiving signal in one symbol as

$$\mathbf{Y}_{VA} = \left[\mathbf{Y}_{1,1}^T,...,\mathbf{Y}_{\eta_x,1}^T,\mathbf{Y}_{1,2}^T,...,\mathbf{Y}_{\eta_x,2}^T,.......,\mathbf{Y}_{\eta_x,\eta_y}^T\right]^T \in \mathbb{C}^{\eta_x\eta_y L_x L_y \times N}.\quad (48)$$

With Equation (48), the VA can now be applied to the OFDM-based JCAS system.

### C. Filter Design

For the wideband chirp signal in MaRS, the filter design is the same as the conventional chirp radar. However, filtering the single-tone signal in MaRS is more challenging than that of the Doppler radar using single-tone signal. It can be discontinuous as the MaRS signal should have a duty ratio lower than 100% . Also, the receive array receives the single-tone sensing signal together with the uplink communication signal.

This paper proposes a 2-step filtering based receiver to cancel both the uplink communication signal and the sensing SI as shown in Fig. 10. There are $M$ pulse repetition intervals (RPI) in one JCAS frame. One RPI contains two parts including the sensing-communication-shared signal and the communication-only signal. The shared signal is processed by a low-pass filter (LPF), which removes the uplink communication signal. Then, a sample and hold (S/H) module is used to integrate the received sensing signal and hold the integral value. During the analog integral, the sum in Equation (26) is done, which also greatly saves the hardware storage. Every value is held for one PRI, and these values form a DC-biased staircase single-tone signal. In the last step, a band-pass filter (BPF) is used to remove the DC bias and staircase edges, and the sing-tone echo with target Doppler information can be extracted.

## V. NUMERICAL RESULTS

### A. Parameter Settings

This paper considers simulation scenarios of both cars and unmanned aerial vehicles (UAV) detection as shown in Fig. 1. The simulation parameters are listed in Table 1. The received SI power $P_{SI}$ is usually lower than the transmit power by 30~72 dB according to a survey paper on full duplex [45]. Here, $P_{SI}$ is set to be 50 dB lower than the transmit power. A CP length of 8.33 us is assumed in simulations, which covers a inter-carrier-interference(ICI)-free sensing range of 1.25 km. If a maximum sensing range is 625 m, the 60kHz sub-carrier spacing extend CP length of 4.17 us can be used. The GTRI empirical model [40] of grass is used to simulate environmental clutters. Since the GTRI model does not contain the distance information, this paper divides a 2 km x 1 km area into $2 \times 10^6$ 1m x 1m grids and uses the grid centers to generate the distances. This step can be generated with unit power for one time and reused in different simulation scenes.

The power allocation strategy of L-shape MaRS and comb MaRS is to allocate the chirp part with 10-fold average transmit power and the single-tone part with $(M-10W)/(M-W)$ of the average transmit power. As a comparison, the wideband chirp signal and FA MaRS directly uses the average transmit power. The parameters of comb MaRS are $W = 4$, $[i_0, i_1, i_2, i_3]$ = [1, 3, 6, 10] for car detection and $[i_0, i_1, i_2, i_3]$ = [1, 6, 16, 30] for UAV detection, which are decided by different velocity ranges. The height of the UAV is 0~100 m. Using the parameters in this table, the sensing range resolution is 1.22 m. The angular resolution $\Delta\sin\psi$ is 1/4 without VA and 1/8 with VA, which corresponds to a resolution of 14.5° and 7.18° at $\psi$ = 0°. The speed resolution is 0.24 m/s, while the maximum ambiguous speed is ± 120 m/s. The estimation accuracy can be beyond the resolution when super resolution method is used. The detection hit rate is defined as the percentage of targets in different drops which are successfully detected in the ground-truth resolution unit or the unit adjacent to it, as the ground-truth value may lie in the middle of two resolution units.

### B. Performance Comparison

Fig. 11 shows the detection hit rate comparison of different sensing signals. Comb MaRS performs best and achieves an over 90% detection hit rate up to 1000 m for both two scenarios. Also, the performance of comb MaRS is close to that of conventional chirp signal which is widely used in commercial radars [44] with the same average transmit power. L-shape MaRS performs worse than comb MaRS. The gap is larger for UAV detection as it is more difficult to separate small-RCS targets from strong clutters. FA MaRS performs badly, and it can only be used for car detection at a very close range. The reason is that FA MaRS gets only the phase of every symbol for both range and Doppler estimation, and the strong static clutter signals are not removed or suppressed before detection.

As the OFDM-based scheme is used, the communication performance can be measured by the number of REs. Fig. 12 shows the sensing overhead percentages of different schemes. Compared with the wideband chirp, the proposed MaRS reduces the sensing overheads by several orders of magnitude, while the performance is still good enough for the two scenarios. For the simulation setting of $M = 500$ without VA, the sensing overheads of the conventional chirp scheme are up to 10%, while that of MaRS is lower than 0.1%. When VA is used, the resource overheads increase by around 3-fold. The low sensing resource overheads without and with VA shows a high communication resource utilization ratio of at least 99.9% and 99.6% in the proposed JCAS schemes.

| Parameter | Value |
|---|---|
| Carrier frequency, $f_C$ | 5 GHz |
| Sub-carrier spacing, $\Delta f$ | 60 kHz |
| Bandwidth, $B$ | 122.88 Mhz |
| Total sub-carrier number, $N$ | 2048 |
| Average sensing TX power | 10 dBm (Car) <br> 45 dBm (UAV) |
| Noise power density | −174 dBm/Hz |
| PRI | 0.25 ms |
| CPI | 125 ms |
| Sensing OFDM symbol number, $M$ | 500 |
| Sensing symbol length | 16.67 us |
| Large URA size, $(L_x, L_y)$ | (8, 8) |
| Small URA size, $(\eta_x, \eta_y)$ | (2, 2) |
| Radar cross section | 100 m² (Car) <br> 0.02 m² (UAV) |
| Sensing range | 100~1000 m |
| Speed range | 5~300 km/h (Car) <br> 5~80 km/h (UAV) |
| Clutter model | GTRI Model @ 5GHz |
| Butterworth filter order | 5 |
| Uplink transmit power | 23 dBm |
| Uplink Node Distance | 100 m |

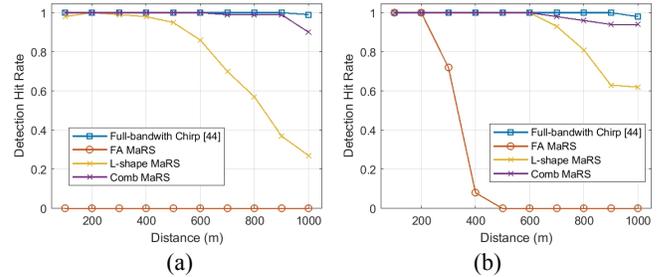

Fig. 11. The detection hit rate of FA MaRS, L-shape MaRS and comb MaRS in (a) car detection and (b) UAV detection. VA is enabled while the super resolution algorithm is not used.

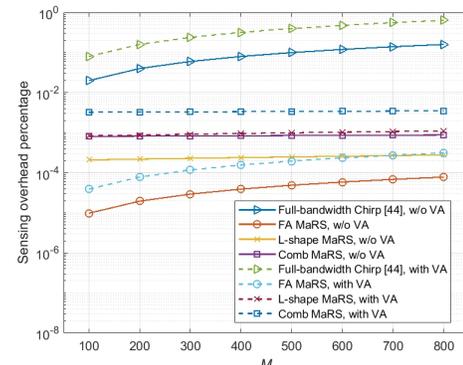

Fig. 12. The sensing resource overheads of different JCAS waveforms in the same scenario.

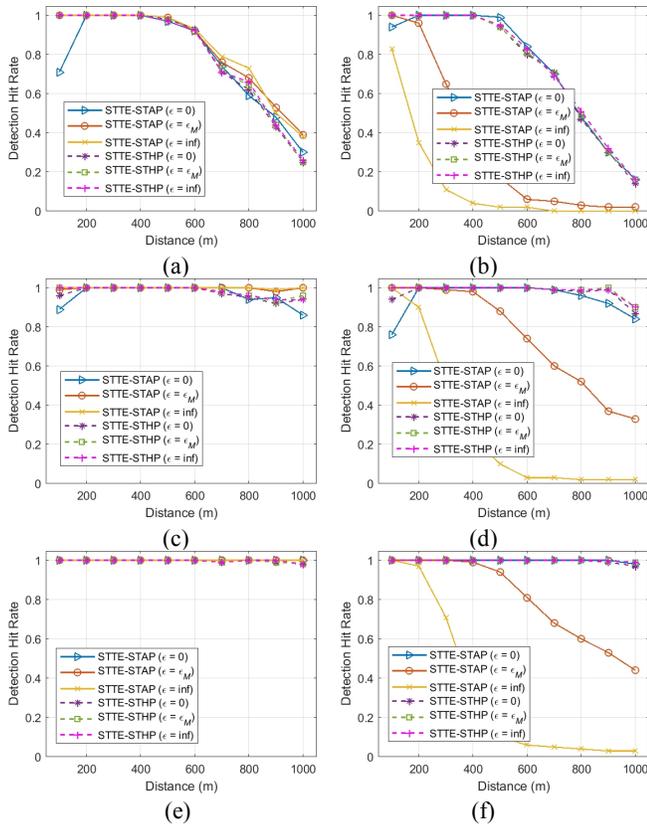

Fig. 13. The single-target detection hit rates of different ADM-based algorithms. (a) and (b) represent the performance of FFT method without VA for car and UAV detection, (c) and (d) represent the performance of FFT method with VA for car and UAV detection, and (e) and (f) represent the performance of MUSIC method with VA for car and UAV detection.

Different ADM-based methods are compared with different distances. In Fig. 13(a), the detection hit rate reduces to under 90% at around 600 m. Also, a performance degradation at 100 m is found for STTE-STAP ($\varepsilon = 0$), which results from $\hat{\mathbf{a}}_{ST,k}/\hat{\mathbf{S}}_1$ mismatch problem mentioned before. In Fig. 13(b), the detection hit rate decreases to under 90% around 500m. The $\hat{\mathbf{a}}_{ST,k}/\hat{\mathbf{S}}_1$ mismatch problem still exists for STTE-STAP ($\varepsilon = 0$). In UAV detection, STTE-STAP with $\varepsilon = \varepsilon_M$ and $\varepsilon \to \infty$ performs much worse due to weaker RCS. Note the clutter cannot be suppressed by higher transmit power, as it also increases with transmit power. Fig. 13(a) and (b) show that STTE-STAP suffers from $\hat{\mathbf{a}}_{ST,k}/\hat{\mathbf{S}}_1$ mismatch problem for $\varepsilon = 0$ and clutter interference for $\varepsilon = \varepsilon_M$ and $\varepsilon \to \infty$, while STTE-STHP avoids these problems.

Fig. 13(c) and (d) show the detection hit rates when VA is enabled. The finer angular resolution unit brought by VA provides a more accurate time-space estimation, which also improves the space-time processing performance a lot. In the car detection scenario, STTE-STAP ($\varepsilon = \varepsilon_M / \varepsilon \to \infty$) performs best, which is close to 100% in all range bins. In UAV detection, STTE-STHP ($\varepsilon = \varepsilon_M / \varepsilon \to \infty$) performs best which is close to 100% in 100~900m. In both two scenarios, STTE-STAP ($\varepsilon = 0$) still suffers from the $\hat{\mathbf{a}}_{ST,k}/\hat{\mathbf{S}}_1$ mismatch problem in the near range, and STTE-STHP ($\varepsilon = 0$) only degrades by a little bit due to this problem. The strong clutter interference limits the UAV detection performance of STTE-STAP ($\varepsilon = \varepsilon_M / \varepsilon \to \infty$). Fig. 13(e) and (f) further employs the MUSIC method to get super-resolution space-time vector estimation. The super-resolution ratio is 10-fold. In both two scenarios, all methods gain an almost 100% detection hit rate, apart from STTE-STAP in UAV detection ($\varepsilon = \varepsilon_M / \varepsilon \to \infty$). The $\hat{\mathbf{a}}_{ST,k}/\hat{\mathbf{S}}_1$ mismatch problem is no longer observed as the estimations get much closer to the ground-truth values in this case.

A sensing case of random distance between 100m and 1000 m is assumed, and multiple targets ($K = 5$) are also considered. Fig. 14 shows the normalized RMSE performance comparison, which normalizes RMSE by the resolution unit without VA. Only the errors of the successful detection are taken into account as the unsuccessful detection leads meaningless and random error values. The successful detection rates can be found in Fig. 15. As mentioned before, $\Delta \sin\psi_x$ and $\Delta \sin\psi_y$ are 1/4, which is equivalent to 14.5° at 0° angle, $\Delta v$ is 0.24 m/s, and $\Delta R$ is 1.22 m. The RMSE caused by the resolution granularity is

$$\int_{-1/2}^{1/2} x^2 dx = 6/\sqrt{3} \approx 0.2887. \quad (49)$$

This value is observed in the normalized RMSE of different parameters using FFT methods without VA.

In the car detection scenario of Fig. 14(a)~(d), VA effectively decreases RMSE of both two angle estimations to one half, and the super-resolution method further increases the RMSE accuracy by around 4-fold. For the velocity estimation, the MUSIC method reduces RMSE to lower than 1/3. For the range estimation, different schemes achieve similar RMSE. MUSIC method is not used for distance estimation as there are only $W$ distance measurement samples. Similar results can be found for UAV detection in Fig. 14 (e)~(h). The results of lower transmit power fluctuate as the corresponding hit rate is low as shown in Fig. 15. One main difference is that MUSIC increases the RMSE performance by more times. Using MUSIC, the RMSE accuracy of $\sin\psi_x$ and $v$ are improved to around 7 and 9 times. The explanation is that UAV has a more free space-distribution as the height is random. The RMSE performance of $\sin\psi_y$ is not increased as the values of $\sin\psi_y$ are still mainly focused on angles around 0° with large distances and limited heights.

For the multiple target detection with random distances in Fig. 15, the detection hit rate increases with the transmit power $P_{tx}$, and it saturates as the performance is limited by SNR-irrelevant factors like environmental clutters and multiple target interference. In the car detection scenario, the STTE-STAP ($\varepsilon = \varepsilon_M$) gains the best performance, and the gap between this method and STTE-STHP ($\varepsilon = \varepsilon_M$) becomes narrower when VA and MUSIC are further used for enhancement. The $\hat{\mathbf{a}}_{ST,k}/\hat{\mathbf{S}}_1$ mismatch problem is also observed for STTE-STAP/STHP ($\varepsilon = 0$) when $P_{TX}$ is high. When VA is not used, STTE-STAP ($\varepsilon = \varepsilon_M$) reaches a saturated value of 86.6% at $P_{tx}$= 30 dBm. When VA is used, it gets to saturation of 95.8% at 20 dBm. When MUSIC is further used, the hit rate performance is increased, and the saturation becomes 96.2% at 10 dBm. For the UAV detection scenario, both STTE-STAP ($\varepsilon = \varepsilon_M$) and STTE-STHP ($\varepsilon = \varepsilon_M$) perform best. The optimal hit rate increases from 74.8% to 85.0% with the VA enhancement. It further increases to 96.2% when MUSIC is further used.

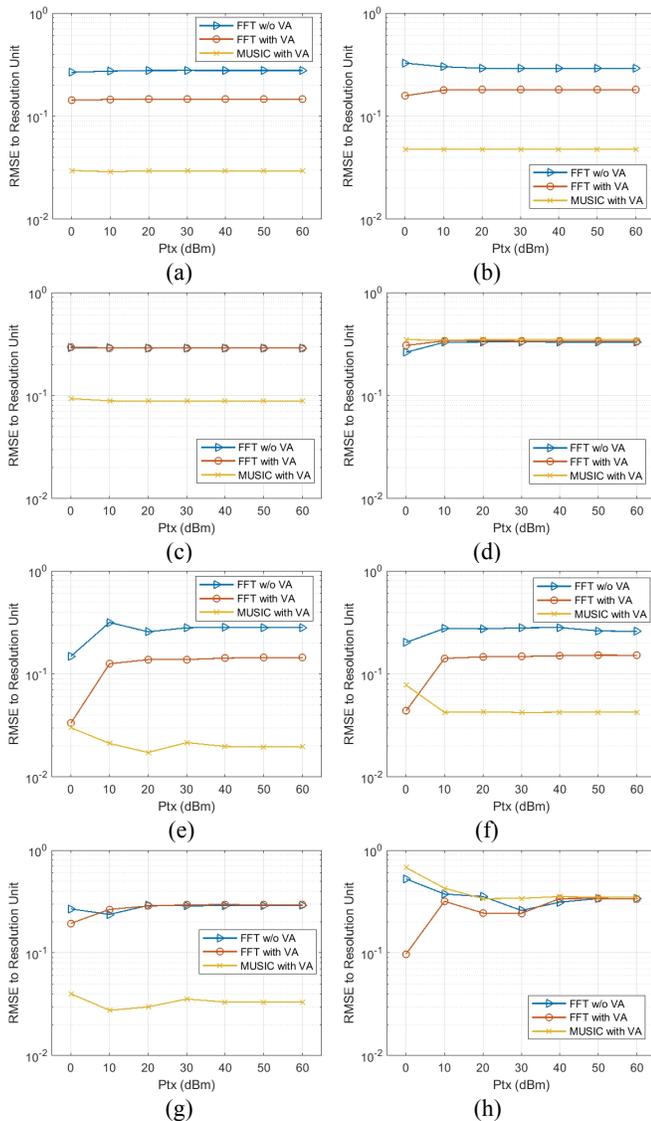

Fig. 14. The normalized RMSE performance comparison. (a)~(d) are the RMSE of $\sin(\psi_x)$, $\sin(\psi_y)$, $v$ and $R$ in car detection scenario, and (e)~(h) are the RMSE of $\sin(\psi_x)$, $\sin(\psi_y)$, $v$ and $R$ in UAV detection scenario. STTE-STHP ($\varepsilon \to \infty$) is used in this simulation.

To verify why 100% hit rate cannot be achieved, a single-user case with random distance is also simulated in Fig. 16, which can gain 100% hit rates in both scenarios. This result most shows that the multi-target interference leads to unachievable 100% hit rate. As the echo power is a quartic inverse proportion to the distance, the power difference among different targets can be large. The worst case in the simulation is one target at 100m and another at 1000m, and the power difference is up to 40 dB.

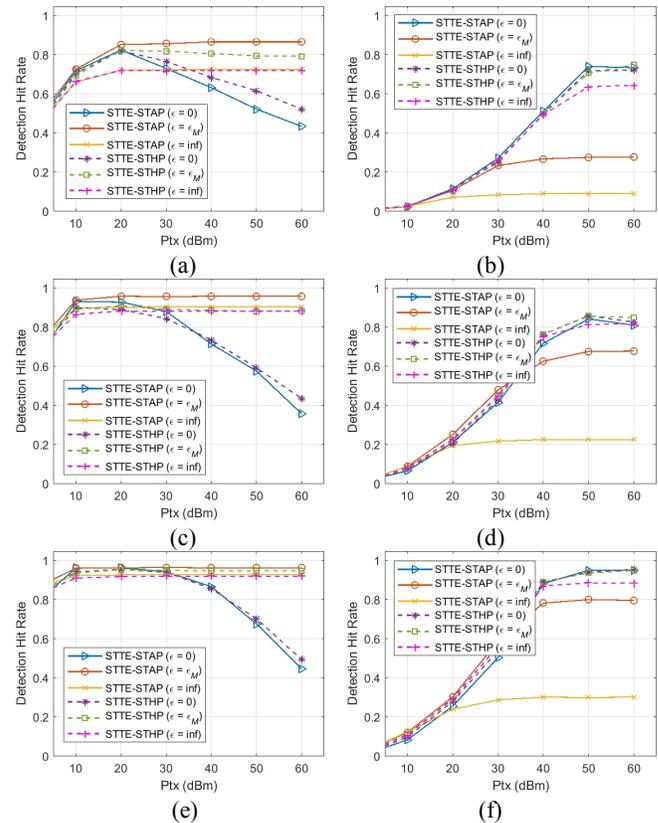

Fig. 15. The random-distance and multi-target detection hit rates of different ADM-based algorithms. (a) and (b) are the performance of FFT method without VA for car and UAV detection, (c) and (d) are the performance of FFT method with VA for car and UAV detection, and (e) and (f) are the performance of MUSIC method with VA for car and UAV detection.

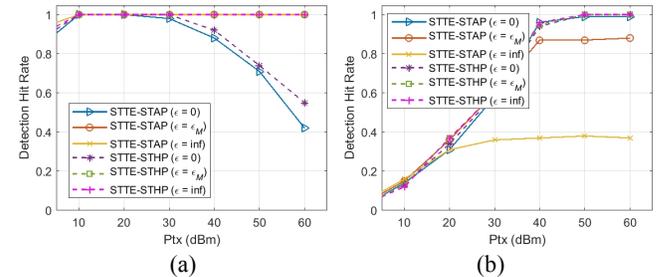

Fig. 16. The random-distance and single-target detection hit rates of MUSIC methods with VA for (a) car and (b) UAV detection.

## VI. CONCLUSIONS

This paper proposes the MaRS waveform design to orchestrate sensing and communication which allows the former to gain a large aperture and the latter to use most of resources. MaRS keeps the radar advantages of constant-modulus, zero auto-correlation and simple SIC. Different time-frequency MaRS structures are proposed and compared, including FA MaRS, L-shape MaRS, and comb MaRS. Comb MaRS performs best as it enables space-time processing after the detection and estimations in ADM. Two ADM-based methods named STTE-STAP and STTE-STHP algorithms are proposed. The latter avoids large-dimension matrix inverse and increases the calculation accuracy when SNR is high. Apart from the waveform and algorithm, the hardware structure of the hybrid duplex is proposed, which avoids the complex IBFD communication. The hybrid duplex only adds a small sensing-dedicated antenna array to the existing half-duplex communication array. The small array is designed with a larger antenna spacing, and two arrays form a large VA in the space domain. The much simpler full duplex of sensing signal is required, which is realized by the analog filter. A 2-step filtering based receiver is also proposed to support the uplink coexistence of radar and communication signals. The numerical results verify the proposed schemes and show that the range, velocity, and angle resolution of a large time-frequency-space aperture can be gained with the very limited resource and hardware overheads.

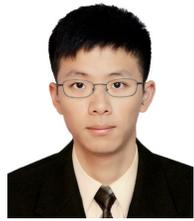

Yihua Ma (yihua.ma@zte.com.cn) received the B.Eng. degree from Southeast University, China in 2015 and the M.Sc degree from Peking University, China in 2018. Since 2018 he is with ZTE Corporation. He is now an expert-level research engineer in the Department of Wireless Algorithm, ZTE, and also a member of State Key Laboratory of Mobile Network and Mobile Multimedia Technology. His main research interests include integrated sensing and communications, mMTC, grant-free transmissions, NOMA and cell-free massive MIMO.

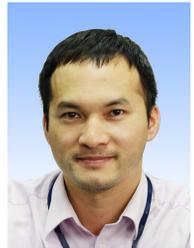

Zhifeng Yuan (yuan.zhifeng@zte.com.cn) received MS degree in signal and information processing from Nanjing University of Post and Telecommunications in 2005. From 2004 to 2006, he was mainly engaged in FPGA / SOC ASIC design. He has been as a member of the wireless technology advance research department at ZTE since 2006 and has been responsible for the research of the new multiple access group since 2012. His research interests include wireless communication, MIMO systems, information theory, multiple access, error control coding, adaptive algorithm, and high-speed VLSI design.

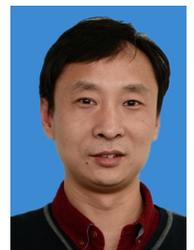

Shuqiang Xia (xia.shuqiang@zte.com.cn) is a senior communication research expert in ZTE. His research interests focus on Integrated Sensing and Communications (ISAC), carrier aggregation (CA), and Ultra-Reliable Low-Latency Communications (URLLC). He received his master's degree in signal and information processing from Nanjing University of Science and Technology, China, in 2002. He was a recipient of the China Patent Gold Award and Second Prize of National Technological Invention Award.

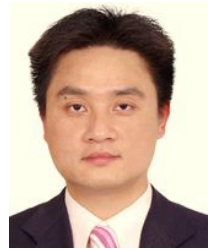

Yu Guanghui (yu.guanghui@zte.com.cn) received the Master and Ph.D. degrees in automatic control from Beijing Institute of Technology, Beijing China, in 1998 and 2003 respectively. Since 2003, he has acted as a radio expert in the wireless advanced research department in ZTE Corporation. His research interests include 3G, 4G, 5G and 6G design in RAN especially involved in multiplexing & access, MIMO, interference management, channel modeling AI/ML, sensing as well as network architecture. He is also involved as one of the main researchers in the link and system simulation platform takes part in all kinds of 3GPP RAN1 activities.

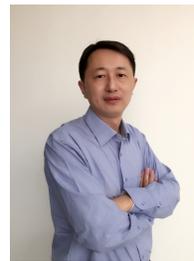

Liujun Hu (hu.liujun@zte.com.cn) received his Doctor's degrees in Electronic and Information from University of Electronic Science and Technology of China. He has more than 20 years of experience in telecom field. He has extensive experience in base band signal processing, mobile network planning and design, with 150 patents and about 15 papers in wireless system, including 3G\4G\5G and future next general moblie system.